\documentclass[journal,12pt,draftclsnofoot,onecolumn]{IEEEtran}
\usepackage{mathrsfs}
\usepackage[cmex10]{amsmath}
\usepackage{amsfonts,amssymb}
\usepackage{mdwmath}
\usepackage{amsthm}
\usepackage{subfigure}
\usepackage{epsf}
\usepackage{graphicx}
\usepackage{subfigure}
\usepackage[sort,compress]{cite}
\usepackage{algorithmic}
\usepackage[ruled,vlined]{algorithm2e}
\usepackage{booktabs} 
\usepackage{multirow}
\usepackage{color}          
\usepackage{threeparttable} 
\usepackage{lipsum,multicol}

\newtheorem{theorem}{Theorem}

\theoremstyle{definition}

\begin{document}
\begin{titlepage}
\begin{center}
\vspace*{-2\baselineskip}
\begin{minipage}[l]{7cm}
\flushleft
\includegraphics[width=2 in]{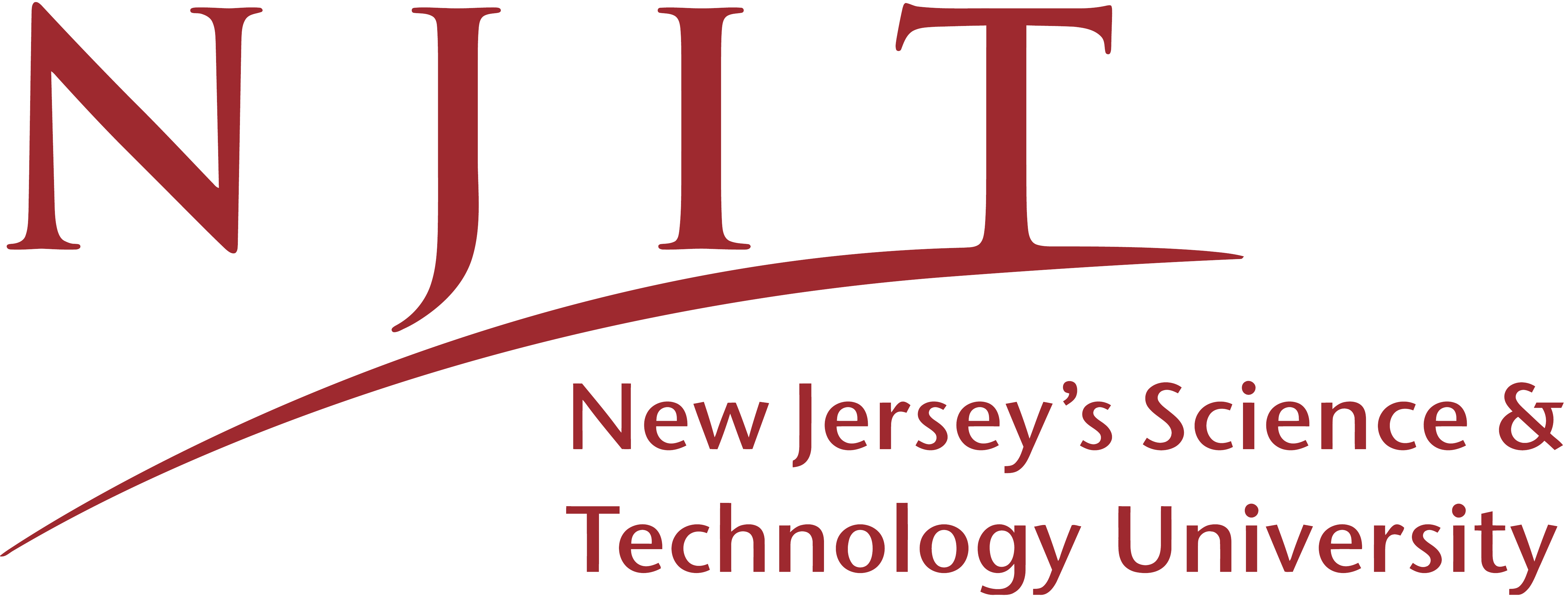}
\end{minipage}
\hfill
\begin{minipage}[r]{7cm}
\flushright
\includegraphics[width=1 in]{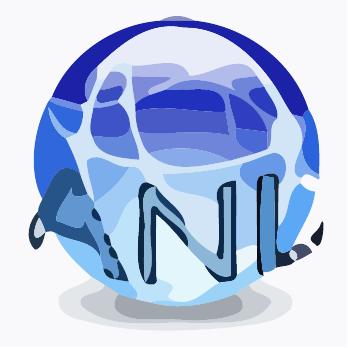}
\end{minipage}
\vfill
\textsc{\LARGE Backhaul-aware Uplink Communications in   \\[12pt]
Full-Duplex DBS-aided HetNets}\\
\vfill
\textsc{\LARGE LIANG ZHANG\\[12pt]
\LARGE NIRWAN ANSARI}\\
\vfill
\textsc{\LARGE TR-ANL-2019-001\\[12pt]
\LARGE July 18, 2019}\\[1.5cm]
\vfill
{ADVANCED NETWORKING LABORATORY\\
 DEPARTMENT OF ELECTRICAL AND COMPUTER ENGINEERING\\
 NEW JERSY INSTITUTE OF TECHNOLOGY}
\vfill
\end{center}

\begin{minipage}[c]{16cm}
\flushleft
\small
{{Citation:}\\
L.~Zhang and N.~Ansari, {``Backhaul-aware uplink communications in full-duplex DBS-aided HetNets,"} to be presented at \emph{IEEE GLOBECOM 2019}, Dec. 9--14, 2019. \\}
\end{minipage}

\end{titlepage}
\title{Backhaul-aware Uplink Communications in Full-Duplex DBS-aided HetNets}
\author{\IEEEauthorblockN{Liang~Zhang, ~\IEEEmembership{Student member,~IEEE},
                          and~Nirwan Ansari, ~\IEEEmembership{Fellow,~IEEE}}
\thanks{
This work was supported in part by U.S. National Science Foundation under Grant CNS-1814748.}}
\maketitle

\begin{abstract}
Drone-mounted base stations (\emph{DBS}s) are promising solutions to provide ubiquitous connections to users and support many applications in the fifth generation of mobile networks while full duplex communications has the potential to improve the spectrum efficiency. In this paper, we have investigated the \underline{b}ackhaul-aware \underline{u}plink communications in a full-duplex \underline{D}BS-aided HetNet (\emph{BUD}) problem with the objective to maximize the total throughput of the network, and this problem is decomposed into two sub-problems: the DBS Placement problem (including the vertical dimension and horizontal dimensions) and the \underline{joint} \underline{U}E association, \underline{p}ower and \underline{b}andwidth assignment (\emph{Joint-UPB}) problem. Since the BUD problem is NP-hard, we propose approximation algorithms to solve the sub-problems and another, named the AA-BUD algorithm, to solve the BUD problem with guaranteed performance. The performance of the AA-BUD algorithm has been demonstrated via extensive simulations, and it is superior to two benchmark algorithms with up to $45.8\%$ throughput improvement.
\end{abstract}
\IEEEpeerreviewmaketitle


\section{Introduction}
The fifth generation of mobile technology (\emph{5G}) targets to provide better performance as compared to 4G LTE, i.e., greater throughput, lower latency and ultra-high reliability \cite{NGMN_5G:2015}. Full duplex (\emph{FD}), which facilitates simultaneous transmission and reception over the same frequency spectra, is a promising technology to improve the spectrum efficiency for the next generation of wireless networks to overcome the shortage of spectrum \cite{NGMN_5G:2015, Spectrum_sharing_2018:WC}. Drone-mounted base stations (\emph{DBS}s) are able to  provide ubiquitous connections to diversified user equipments (\emph{UE}s) because of their flexibility, and efficient and high quality-of-service (\emph{QoS}) provisioning, especially useful for supporting unexpected and temporary events \cite{Yaliniz:2016:NFR, Liang_DBS:2019:IEEE_WC}.

Many works related to DBS communications, viz., Unmanned Aerial Vehicle Base Station (\emph{UAV-BS}) communications~\cite{3D_UAV_placment:2018IEEE_WCL, 3D_placement_ICC_2016, Placement_UAV:2017:IEEE_CL}, have been reported. Alzenad \emph{et al.}~\cite{3D_UAV_placment:2018IEEE_WCL} studied the UAV-BS placement problem with the target to maximize the number of served UEs, and they proposed an exhaustive search algorithm to obtain the the optimal altitude and coverage radius under a given path loss threshold. Bor-Yaliniz \emph{et al.}~\cite{3D_placement_ICC_2016} highlighted the properties of the 3-D DBS placement problem with the objective to maximize the revenue, which is proportional to the number of covered UEs. Lyu \emph{et al.}~\cite{Placement_UAV:2017:IEEE_CL} investigated the UAV-BS placement problem, and the objective is to minimize the number of required DBSs while each UE is at least covered by one DBS. There are also many works about FD communications~\cite{FD_OFDMA:2015TWC, FD_Relay_fading:2016:IEEE_ICC, Joint_FD:2018:IEEE_TVT}. Nam \emph{et al.}~\cite{FD_OFDMA:2015TWC} maximized the total throughput of all FD-enabled UEs in an FD orthogonal frequency division multiple access (\emph{OFDMA}) network with only one BS. Goyal \emph{et al.}~\cite{FD_Relay_fading:2016:IEEE_ICC} studied the spectral efficiency of a mixed multi-cell network, viz., mixed FD and HD cells while all UEs are half-duplex (\emph{HD}) enabled. Chen \emph{et al.}~\cite{Joint_FD:2018:IEEE_TVT} maximized the total sum-rate of uplink and downlink communications within one FD BS under a heavy workload scenario.


Few works have addressed the uplink communications in the HetNet with IBFD enabled DBSs. In our previous work~\cite{Liang_DBS:2018_CL}, we investigated the throughput maximization of the downlink communications in a HetNet with in-band full-duplex (\emph{IBFD}) enabled DBSs. In this paper, we study the \underline{b}ackhaul-aware \underline{u}plink communications in a full-duplex \underline{D}BS-aided HetNet (\emph{BUD}) problem.

The main contributions of this paper are delineated as follows: 1) we have proposed an IBFD-enabled DBS-aided HetNet for uplink communications, and the DBSs can provide dynamic coverage to UEs by adjusting their vertical positions and horizontal positions; 2) the macro-BS (\emph{MBS}) is connected to the core network through free space optics (\emph{FSO}) links, implying that this network can be easily deployed to provide communications to temporary events or fast communications recovery in emergency situations; 3) we propose two approximation algorithms to solve the sub-problems and another one named AA-BUD algorithm to solve the BUD problem. The AA-BUD algorithm with the approximation ratio of $\frac{1}{2}$ is shown capable of acquiring the optimal locations of all DBSs.

The remainder of this paper is organized as follows. The system model in described in Section \ref{sec:system-model} and the BUD problem is formulated in Section \ref{sec:problem-formulation}. Then, two approximation algorithms are proposed to solve the sub-problems and another one named AA-BUD algorithm is proposed to solve the BUD problem in Section \ref{sec:analysis}. Section \ref{sec:evaluation} presents the performance of the AA-BUD algorithm and the comparison with two benchmark algorithms. Finally, conclusions are drawn in Section \ref{sec:conclusion}.

\begin{figure}[!htb]
    \centering
    \includegraphics[width=1.0\columnwidth]{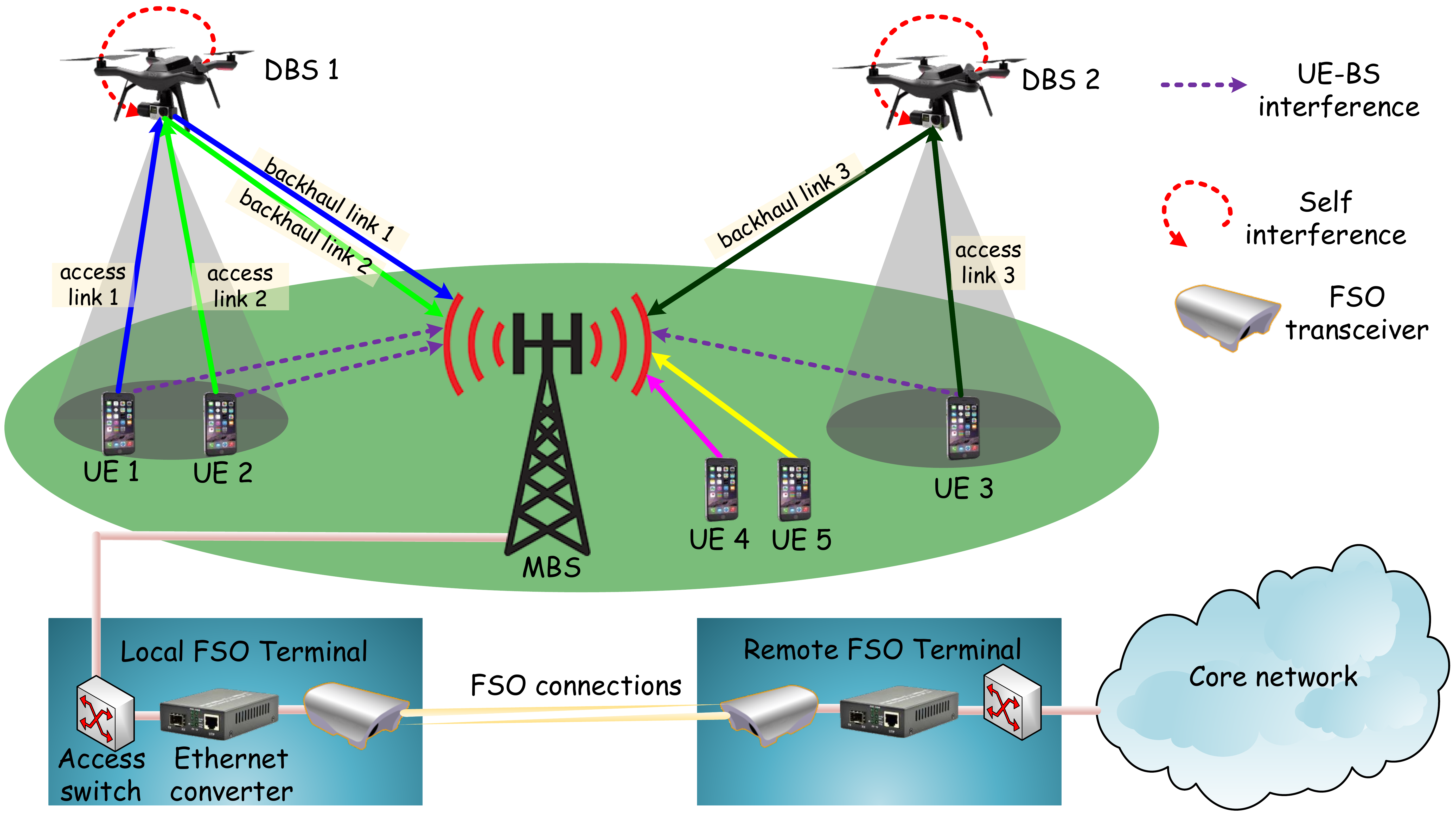}
    \caption{\small The IBFD DBS-aided HetNet framework.}
    \label{fig:full_duplex}
\vspace{-2mm}
\end{figure}

\section{System Model}\label{sec:system-model}
Fig.~\ref{fig:full_duplex} shows a DBS-aided HetNet, in which the frequency division duplex (\emph{FDD}) OFDMA framework is adopted~\cite{FD_OFDMA:2015TWC}. DBS $1$ and DBS $2$ are FD-enabled, and the MBS and all UEs are HD-enabled. The MBS is connected to the core network through the local FSO terminal and the remote FSO terminal. Both the local FSO terminal and the remote FSO terminal include an access switch, an Ethernet converter (Ethernet/FSO signal conversion) and an FSO transceiver. The distance between the local FSO terminal and the remote FSO terminal can be a few kilometers while a high data rate transmission can still be achieved \cite{FSO_backhaul:2017:Mobicom}. For example, Sarkar \emph{et al.}~\cite{FSO_120Gbps_1km:2018} designed a 64-QAM FSO transceiver for one hop transmission, and the transceiver demonstrates a $120$ $Gb/s$ reliable communication data rate over a $1$ $km$ link. An access link is the link from a UE to a BS (DBS), and a backhaul link is the link from a DBS to the MBS.

As shown in Fig.~\ref{fig:full_duplex}, different UEs utilize different frequency spectra for communication, no matter whether the UEs are associated with the DBS (UE 1 and UE 2) or the MBS (UE 4 and UE 5); different DBSs are assigned with different frequency spectra (UE 1, UE2 and UE 3); the backhaul link of a DBS reuses the frequency spectra of its access link (access link 1 and backhaul link 1). In this work, we focus on the uplink communications. In other words, we focus on data transmission from a UE to the MBS directly or via a FD-enabled DBS. For the uplink communications, the basic (minimum) unit of the frequency spectrum is one subcarrier (\emph{SC}); one UE can be provisioned by one or multiple SCs while one subcarrier can only be assigned to one UE in order to avoid UE-UE interference.

\subsection{Path Loss Model}
For the path loss of the proposed framework in Fig.~\ref{fig:full_duplex}, we consider air-to-ground (\emph{A2G}) path loss (DBS-MBS) and ground-to-air (\emph{G2A}) path loss (UE-DBS). For both A2G and G2A path loss, we consider line-of-sight (\emph{LoS}) and none-line-of-sight (\emph{NLoS}) path loss~\cite{Optimal_LAP_Altitude2014, Liang_DBS:2018_CL, Liang_DBS:2019:IEEE_TVT}. Denote $\psi_{i, j}^{L}$ and $\psi_{i, j}^{N}$ as the probability of a LoS and NLoS connection of an A2G (G2A) link, as shown in Eq.~\eqref{eq:em1}. Here, $a$ and $b$ are environment constants (i.e., suburban, urban or dense urban); $\theta_{i, j}=arctan(\frac{h_{j}}{d_{i,j}})$ is the elevation angle; $h_{j}$ ($j>1$) is the altitude of the $j$th DBS and $d_{i,j}$ ($j>1$) is the 3-D distance between the $i$th UE and the $j$th DBS~\cite{3D_UAV_placment:2018IEEE_WCL, Optimal_LAP_Altitude2014}.

\vspace{-2mm}
\begin{equation}\label{eq:em1}
\left\{
\begin{aligned}
&\psi_{i, j}^{L}=[1+a \cdot exp(-b(\frac{180\theta_{i, j}}{\pi}-a))]^{-1},\\
&\psi_{i, j}^{N}=1-\psi_{i, j}^{L}.
\end{aligned}
\right.
\end{equation}

Let $\eta_{i,j}$ be the path loss between the $i$th UE and the $j$th DBS, as described in Eq.~\eqref{eq:em2}. Here, $\zeta^{L}$ and $\zeta^{N}$ are the additional path loss of LoS and NLoS, respectively; $f_{0}$ is the carrier frequency and $c_{0}$ is the transmission speed of light. The first item is the excessive path loss of LoS, the second item is the excessive path loss of NLoS, and the third item is the mean free-space path loss (including LoS and NLoS free-space path loss).
\vspace{-2mm}
\begin{equation}\label{eq:em2}
\eta_{i,j}= \psi_{i, j}^{L}\zeta^{L}+\psi_{i, j}^{N}\zeta^{N} + 20log(4\pi f_{0}d_{i, j}/c_{0}).
\end{equation}

\vspace{-2mm}
After substituting Eq.~\eqref{eq:em1} into Eq.~\eqref{eq:em2}, we have
\begin{equation}\label{eq:em3}
\eta_{i,j}= \psi_{i, j}^{L}(\zeta^{L}-\zeta^{N}) +20log(4\pi f_{0}d_{i, j}/c_{0})+\zeta^{N}.
\end{equation}

\vspace{-2mm}
\subsection{Communications Model}
 Let $s_{i, j}^{1}$ and $s_{i, j}^{2}$ ($j>1$) be the signal to interference plus noise ratio (\emph{SINR}) of the access link and the backhaul link from the $i$th UE to the MBS via the $j$th DBS, as expressed in Eqs.~\eqref{eq:em4}-\eqref{eq:em5}. Here, $j=1$ implies that the UE connects to the MBS directly; $P_{U}$ is the transmission power of a UE; $\sigma_{j}^{2}\!=\!\tau_{0}b_{i, j}N_{0}$ is the thermal noise power, $\tau_{0}$ is the bandwidth of one SC, $b_{i, j}$ is the assigned bandwidth for the $i$th UE to the $j$th BS in terms of SCs, and $N_{0}$ is the thermal noise power spectral density; $\alpha_{i,j}=p_{i, j}/\tau^{SI}$ is the self interference (\emph{SI}) at the $j$th DBS incurred by the FD communications, $p_{i, j}$ is the assigned power by the $j$th DBS for the backhaul link (the $j$th DBS to the MBS) in provisioning the $i$th UE, and $\tau^{SI}$ is the SI cancellation capability \cite{FD_NOMA:2017:IEEE:JSAC}.

\begin{equation}\label{eq:em4}
s_{i, j}^{1} =
\begin{cases}
\frac{P_{U}\Gamma_{i, j}}{\sigma_{i, j}^{2}},\quad\forall i \in \mathcal{U}, j \in \mathscr{B}, j=1,  \\
\frac{P_{U} \eta_{i, j}}{\alpha_{i,j}+ \sigma_{i, j}^{2}},\quad\forall i \in \mathcal{U}, j \in \widetilde{\mathscr{B}}.
\end{cases}
\end{equation}

For Eq.~\eqref{eq:em5}, $\tilde{\eta}_{1, j}$ is the channel gain from the $j$th DBS to the MBS; $\Gamma_{i, 1}$ is the channel gain from the $i$th UE to the MBS; $\sigma_{i, 1}^{2}$ is the thermal noise power at the MBS owing to the transmission of the $i$th UE.
\begin{equation}\label{eq:em5}
s_{i, j}^{2} =\frac{p_{i, j}\tilde{\eta}_{1, j}}{P_{U}\Gamma_{i, 1}+ \sigma_{i, 1}^{2}},\quad\forall i \in \mathcal{U}, j \in \widetilde{\mathscr{B}}.
\end{equation}

Let $\beta_{i, j}$ be the data rate of the $i$th UE towards the $j$th BS. Then, $\beta_{i, j}$ can be calculated by Eq.~\eqref{eq:em6}. Here, $\beta_{i, j}^{1}$ is the data rate of the access link (UE-BS) and $\beta_{i, j}^{2}$ is the data rate of the backhaul link (DBS-BS), as expressed in Eq.~\eqref{eq:em7}.

\vspace{-2mm}
\begin{equation}\label{eq:em6}
\beta_{i, j} =
\begin{cases}
\beta_{i, j}^{1},\quad\forall i \in \mathcal{U}, j \in \mathscr{B}, j=1,  \\
\min (\beta_{i, j}^{1}, \beta_{i, j}^{2}),\quad\forall i \in \mathcal{U}, j \in \widetilde{\mathscr{B}}.
\end{cases}
\end{equation}
\vspace{-2mm}
\begin{equation}\label{eq:em7}
\begin{cases}
\beta_{i, j}^{1}=\tau_{0}b_{i, j}log_{2}(1+s_{i, j}^{1}), \quad\forall i \in \mathcal{U}, j \in \mathscr{B},  \\
\beta_{i, j}^{2}=\tau_{0}b_{i, j}log_{2}(1+s_{i, j}^{2})], \quad\forall i \in \mathcal{U}, j \in \widetilde{\mathscr{B}}.
\end{cases}
\end{equation}

\vspace{-2mm}
\section{Problem formulation}\label{sec:problem-formulation}
We focus on uplink communications in an FD DBS-aided HetNet, and each UE is provisioned by one BS. Notations and variables are listed in Table~\ref{tab:notations}.

\begin{table}[!htb]
\centering
\footnotesize
\begin{center}
\caption{Notations and Variables}\label{tab:notations}
\begin{tabular}{{|l|p{188pt}|}}
\hline
Symbol                  & Definition                                    \\
\hline
$\mathscr{B}$           & The set of BSs (including the MDBS and DBSs).   \\
\hline
$\widetilde{\mathscr{B}}$ & The set of DBSs.                              \\
\hline
$\mathscr{U}$           & The set of UEs.                          \\
\hline
$\mathscr{V}_{1}$       & The set of horizontal candidate locations.    \\
\hline
$\mathscr{V}_{2}$       & The set of vertical candidate locations.       \\
\hline
$\tau_{0}$              & The bandwidth of one SC.                \\
\hline
$r_{i}$                 & The data rate requirement of the $i$th UE.         \\
\hline
$f^{max}$               & The total available bandwidth of all BSs in terms of SCs.\\
\hline
$f_{j}^{max}$           & The total available bandwidth for the $j$th BS in term of SCs.\\
\hline
$P_{D}$                 & The power capacity of the $j$th BS.\\
\hline
$P_{U}$                 & The power capacity of the $i$th UE.\\
\hline
$\kappa_{j}$            & The power spectral density of the $j$th DBS, $j \in \widetilde{\mathscr{B}}$.\\
\hline
$d_{i,j}$               & The 3-D distance between the $i$th UE and the $j$th DBS.\\
\hline
$\eta_{i,j}$            & The path loss between the $i$th UE and the $j$th DBS.\\
\hline
$\tau_{i,j}^{SI}$       & The SI power at the $j$th DBS for provisioning the $i$th UE.\\
\hline
$x_{i, j}$         & The UE-BS association indicator. \\
\hline
$\beta_{i, j}$          & The achieved data rate of the $i$th UE towards the $j$th BS.\\
\hline
$b_{i, j}$              & The assigned SCs by the $j$th BS towards the $i$th UE. \\
\hline
$p_{i, j}$              & The assigned power by the $j$th DBS for the DBS-MBS transmission (backhaul data transmission for the $i$th UE).\\
\hline
$\gamma_{j}$            & The horizontal position of the $j$th BS, $\gamma_{j} \in \mathscr{V}_{1}$.\\
\hline
$h_{j}$                 & The vertical position of the $j$th BS, $h_{j} \in \mathscr{V}_{2}$.\\
\hline
\end{tabular}
\end{center}
\vspace{-4mm}
\end{table}

The BUD problem is formulated as follows. The objective is to maximize the total throughput of the network for the uplink communications, as expressed in Eq.~\eqref{eq:e8}. C1 and C7 are the UE provisioning constraints, which impose one UE to be provisioned by at most one BS. C2 is the bandwidth capacity constraint for each BS and imposes the assigned bandwidth by a BS to its associated UEs not to exceed the BS' bandwidth capacity. C3 is the power capacity constraint of each DBS for the backhaul link, and it imposes the total power used by a DBS not to exceed its power capacity. C4 is the data rate requirement constraint of each UE, implying that the achieved data rate of a UE is equal or larger than the required data rate. C5--C6 are the DBS placement constraints, and they impose all DBSs to be placed on the candidate dimensions in the horizontal plane and vertical plane.

\begin{align}
\mathscr{P}_{0}:& \quad\max_{x_{i, j}, p_{i, j}, b_{i, j},\gamma_{j}, h_{j}}\quad \sum_{i}\sum_{j}{x_{i, j} r_{i}}  \nonumber\\
s.t.:\nonumber & \\
&C1:\sum_{j} x_{i, j} \leq 1,\quad\forall i \in \mathscr{U},                                                 \nonumber\\
&C2: \sum_{i}x_{i, j} b_{i, j} \leq f_{j}^{max},\quad\forall j \in \mathscr{B},                  \nonumber\\
&C3: \sum_{i}{x_{i, j} p_{i, j}} \leq P_{D}, \quad\forall j \in \widetilde{\mathscr{B}},                     \nonumber\\
&C4: x_{i, j}r_{i} \leq \beta_{i, j},\quad\forall i \in \mathscr{U}, j \in \mathscr{B},                   \nonumber\\
&C5: \gamma_{j} \in \mathscr{V}_{1},\quad\forall j \in \widetilde{\mathscr{B}},                                   \nonumber\\
&C6: h_{j} \in \mathscr{V}_{2}, \quad\forall j \in \widetilde{\mathscr{B}},                                        \nonumber\\
&C7: x_{i, j} \in \{0, 1\},  \quad\forall i \in \mathscr{U}, j \in \mathscr{B}.                              \label{eq:e8}
\end{align}

\section{Problem Analysis}\label{sec:analysis}
 Any instance of the Max-Generalized Assignment Problem (\emph{Max-GAP}) problem ~\cite{Max_GAP_MIT:2006_SODA} can be reduced into the BUD problem, and the Max-GAP problem is a well-known NP-hard problem. Thus, the BUD problem is NP-hard. So, we propose to decompose the BUD problem into two sub-problems: the DBS placement problem and the \underline{joint} \underline{U}E association, \underline{p}ower and \underline{b}andwidth assignment (\emph{Joint-UPB}) problem. We first solve the sub-problems one by one, and then we solve the BUD problem.

\subsection{Solving the Joint-UPB Problem}\label{joint-UPB-problem}
For given vertical positions and horizontal positions \cite{location_USTEC:2019} of all DBSs, i.e., $\hat{\gamma}_{j}$ and $\hat{h}_{j}$, $\mathscr{P}_{0}$ can be transformed into $\mathscr{P}_{1}$. Let $\Phi (x_{i, j}, p_{i, j}, b_{i, j}, \gamma_{j}, h_{j}) = \sum_{i}\sum_{j}{x_{i, j}r_{i}}$ be the objective function of $\mathscr{P}_{0}$, and $\left. \Phi_{1}(x_{i, j}, p_{i, j}, b_{i, j})= \Phi \right |_{\gamma_{j}=\hat{\gamma}_{j}, h_{j}=\hat{h}_{j}}$ be the objective function of $\mathscr{P}_{1}$.

\vspace{-4mm}
\begin{align}
\mathscr{P}_{1}:& \quad\max_{x_{i, j}, p_{i, j}, b_{i, j}}\quad \sum_{i}\sum_{j}{x_{i, j} r_{i}}          \nonumber\\
s.t.:\nonumber & \\
&C1, C2, C3, C4, C7 \quad in\quad \mathscr{P}_{0}                                                        \label{eq:e9}
\end{align}

To ensure analytical tractability, we assume the power assignment is proportional to the bandwidth assignment, viz.,  $p_{i,j}=b_{i, j}\kappa_{j}$. Note that the MBS does not assign power and bandwidth to the UEs while the DBSs need to assign power and bandwidth to the backhaul links. Then, constraint $C3$ is relaxed. The required bandwidth to provision the $i$th UE by the $j$th BS can be calculated as $\hat{b}_{i, j}=\operatorname*{argmin}\limits_{b_{i, j}} {(\beta_{i, j}- x_{i, j} r_{i} \geq 0)}, x_{i, j}=1, \forall i \in \mathscr{U}, j \in \mathscr{B}$. Obviously, constraint $C4$ is also relaxed. Then, $\mathscr{P}_{1}$ can be transformed into $\mathscr{P}_{2}$.

\vspace{-4mm}
\begin{align}
\mathscr{P}_{2}:& \quad\max_{x_{i, j}}\quad \sum_{i}\sum_{j}{x_{i, j} r_{i}}                                 \nonumber\\
s.t.:\nonumber & \\
&C1:\sum_{j} x_{i, j} \leq 1,\quad\forall i \in \mathscr{U},                                                 \nonumber\\
&C2: \sum_{i} x_{i, j} b_{i, j} \leq f_{j}^{max},\quad\forall j \in \mathscr{B},                             \nonumber\\
&C3: x_{i, j} \in \{0, 1\},  \quad\forall i \in \mathscr{U}, j \in \mathscr{B}.                              \label{eq:e10}
\end{align}

We propose an approximation algorithm to solve problem $\mathscr{P}_{2}$ as depicted in Algorithm~\ref{AA-UPB}, referred to as Approximation Algorithm for the joint-UPB problem (\emph{AA-UPB}). The parameters are initialized in \emph{Step} $1$. Let the weight be $z_{i}=r_{i}/b_{i, j}$, as calculated by \emph{Steps} $2-6$. Then, the weight is sorted in a decreasing order in \emph{Step} $7$ and this new order represents a new UE sequence. One solution of UE association set $\Lambda_{1}=\cup \{x_{\tilde{i}, \tilde{j}}\}$ is obtained through \emph{Steps} $9-19$. The other solution of UE association set $\Lambda_{3}$, which contains the maximum $|\mathscr{B}|$ UEs, is achieved through \emph{Steps} $20-23$. Finally, the UE association set (either $\Lambda_{1}$ or $\Lambda_{3}$) which produces a higher throughput is returned, and the corresponding $\tilde{b}_{i, j}$ and $\tilde{p}_{i, j}$ are also returned.

\begin{algorithm}[!htb]
\caption{Approximation Algorithm for the joint-UPB problem (\emph{AA-UPB})} \label{AA-UPB}
\small
\SetKwData{Left}{left}\SetKwData{This}{this}\SetKwData{Up}{up}
\SetKwFunction{Union}{Union}\SetKwFunction{FindCompress}{FindCompress}
\SetKwInOut{Input}{Input}\SetKwInOut{Output}{Output}
\Input{$\mathscr{B}$, $\mathscr{U}$, $f_{j}^{max}$, $\kappa_{j}$, $r_{i}$, $\hat{\gamma}_{j}$ and $\hat{h}_{j}$\;}
\Output{$\tilde{x}_{i, j}$, $\tilde{b}_{i, j}$ and $\tilde{p}_{i, j}$\;}
\nl  $\tilde{i}=1$, $f_{j}^{used}=0$, $\Lambda_{0}=\mathscr{U}$, $\Lambda_{1}=\varnothing$, $\forall j \in \mathscr{B}$\;
\nl \For{$i \in \Lambda_{0}$}{
\nl \For{$j \in \mathcal{B}$}{
\nl  $\hat{b}_{i, j}=\operatorname*{argmin}\limits_{b_{i, j}} {(\beta_{i, j}- r_{i} \geq 0)}, \forall i \in \mathscr{U}, j \in \mathscr{B}$\;
\nl  $\hat{p}_{i, j}=\hat{b}_{i, j}\kappa_{j}$\;}
\nl  obtain $\tilde{j}=\operatorname*{argmin}\limits_{j} {\hat{b}_{i, j}}$, $\forall i$\;
\nl  get $b_{i, \tilde{j}}=\min(\widehat{b}_{i, j})$ and $z_{i}=r_{i}/b_{i, \tilde{j}}$\;}
\nl  put the UEs in a descending order $\tilde{i}$ by $z_{i}$\;
\nl  $\Lambda_{2}=\Lambda_{0}$\;
\nl \While{$f_{j}^{used} \leq f_{j}^{max} \& \Lambda_{2} \neq \varnothing$}{
\nl  \If{$f_{j}^{used} + b_{\tilde{i}, \tilde{j}} \leq f_{j}^{max}$}{
\nl       $x_{\tilde{i}, \tilde{j}}=1$\;
\nl       $f_{j}^{used}=f_{j}^{used} + b_{\tilde{i}, \tilde{j}}$\;
\nl       $\Lambda_{1}=\Lambda_{1} \cup \{x_{\tilde{i}, \tilde{j}}\}$\;
\nl       $\Lambda_{2}=\Lambda_{2}\setminus \tilde{i}$\;}
\nl  \Else{
\nl  $\Lambda_{0}= \Lambda_{2}$\;
\nl   go to step 2\;}
\nl  $\tilde{i}=\tilde{i}+1$\;}
\nl $\hat{i}=1$, $\Lambda_{3}=\varnothing$, $\Lambda_{4}=\mathscr{U}$\;
\nl \For{$\hat{i} \leq |\mathcal{B}|$}{
\nl  $\Lambda_{3}=\Lambda_{3} \cup \{\hat{x}_{\hat{i}, \hat{j}}=\operatorname*{argmax}\limits_{x_{i, j}} {x_{i, j} r_{i}}\}$, $\forall i \in \Lambda_{4}$\;
\nl  $\Lambda_{4}=\Lambda_{4}\setminus \hat{i}$\;}
\nl  return $\Lambda_{1}$ or $\Lambda_{3}$ which produces a higher throughput\;
\nl obtain $\tilde{b}_{i, j}$ and $\tilde{p}_{i, j}$.
\end{algorithm}

\vspace{-4mm}
\begin{align}
\mathscr{P}_{3}:& \quad\max_{x_{i, j}}\quad \sum_{i}\sum_{j}{x_{i, j} r_{i}}                                 \nonumber\\
s.t.:\nonumber & \\
&C1, C2 \quad in\quad \mathscr{P}_{2}                                                                        \nonumber\\
&C3: 0 \leq x_{i, j} \leq 1,  \quad\forall i \in \mathscr{U}, j \in \mathscr{B}.                              \label{eq:e11}
\end{align}

\vspace{-2mm}
\begin{theorem}
The AA-UPB algorithm is a $\frac{1}{2}$-approximation algorithm of the problem $\mathscr{P}_{2}$. Especially, this algorithm achieves the optimal throughput when all UEs are provisioned.
\end{theorem}\label{th1}

\vspace{-2mm}
\begin{proof}
Note that problem $\mathscr{P}_{2}$ can be transformed into problem $\mathscr{P}_{3}$ while $x_{i, j}$ is relaxed to a continuous variable. In order to prove \emph{Theorem} $1$, we define $\left. \Phi_{2}(x_{i, j})= \Phi_{1} \right |_{p_{i,j}=p_{\tilde{i}, \tilde{j}}, b_{i,j}=p_{\tilde{i}, \tilde{j}}}$ as the objective function of $\mathscr{P}_{2}$ and $\Phi_{3}(\bar{x}_{i, j})$ be the objective function of $\mathscr{P}_{3}$.

1) If all UEs are provisioned, the achieved total throughput of Algorithm~\ref{AA-UPB} is $\max(\Phi_{2}(\tilde{x}_{\tilde{i}, \tilde{j}}), \Phi_{2}(\hat{x}_{\hat{i}, \hat{j}})) \\= \Phi_{2}(\tilde{x}_{\tilde{i}, \tilde{j}}) = \sum_{i}\sum_{j}{x_{i, j} r_{i}} = \sum_{i} (\sum_{j}x_{i, j}) r_{i}=\sum_{i} r_{i}$; the optimal solutions of $\mathscr{P}_{2}$ and $\mathscr{P}_{3}$ are $\Phi_{2}(x_{i, j}^{*})= \sum_{i}\sum_{j}{x_{i, j}^{*} r_{i}}= \sum_{i} r_{i}$ and $\Phi_{3}(\bar{x}_{i, j}^{*})= \sum_{i}\sum_{j}{\bar{x}_{i, j}^{*} r_{i}} = \sum_{i} r_{i}$, respectively. Here, $\sum_{j}x_{i, j}=1$, $\sum_{j}x_{i, j}^{*}=1$, $\sum_{j} \bar{x}_{i, j}^{*}=1$, $\tilde{x}_{\tilde{i}, \tilde{j}}\in \Lambda_{1}$ and $\hat{x}_{\hat{i}, \hat{j}}\in \Lambda_{3}$. Algorithm~\ref{AA-UPB} produces the results equivalent to the optimal solutions of problem $\mathscr{P}_{2}$ and $\mathscr{P}_{3}$.

2) Here, we discuss the scenario with one or more blocked UEs. We first find the relationship between the optimal value of problem $\mathscr{P}_{3}$, $\Phi_{2}(\Lambda_{1})$ and $\Phi_{2}(\Lambda_{3})$. Then, the lower bound of $\max(\Phi_{2}(\tilde{x}_{\tilde{i}, \tilde{j}}), \Phi_{2}(\hat{x}_{\hat{i}, \hat{j}}))$ is determined, which is leveraged to prove the approximation ratio of the AA-UPB algorithm. Note that Algorithm~\ref{AA-UPB} puts all UEs in a sequence by the decreasing order of the weight (defined by the data rate over the required bandwidth to provision a UE), and all UEs are provisioned by this order until the rest of UEs cannot be served by any BS. Let $(y-1)$ be the index of the last UE which is provisioned in $\Lambda_{1}$, i.e., $|\Lambda_{1}|= y-1$. $\Phi_{3}(\bar{x}_{i, j}^{*}) = \Phi_{2}(\cup_{\tilde{i}=1}^{y-1}{\tilde{x}_{\tilde{i}, \tilde{j}}}) + \epsilon_{\hat{i}} \Phi_{2}(\cup_{\hat{i}=y}^{y-1+|\mathcal{B}|} {\hat{x}_{\hat{i}, \hat{j}}^{'}})$. Here, $\Lambda_{1}=\cup_{\tilde{i}=1}^{y-1}{\tilde{x}_{\tilde{i}, \tilde{j}}}$, and $\cup_{\tilde{i}=y}^{y-1+|\mathcal{B}|} {\hat{x}_{\tilde{i}, \tilde{j}}^{'}}$ includes $|\mathscr{B}|$ UEs with the maximum data rate requirement among the UEs with the starting index $y$ and the end index $|\mathscr{U}|$; $\epsilon_{\hat{i}}=(f_{j}^{max}-\sum_{\tilde{i}=1}^{\tilde{i=y-1}}{\tilde{x}_{\tilde{i}, \tilde{j}}\tilde{b}_{\tilde{i}, \tilde{j}}}) / (\hat{x}_{\hat{i}, \hat{j}}^{'} \hat{b}_{\hat{i}, \hat{j}}^{'})$, $0 \leq \epsilon_{\hat{i}} <1$ and $\hat{x}_{\hat{i}, \hat{j}}^{'}=1$. Note that $\Lambda_{3}=\cup _{\hat{i}=1}^{|\mathscr{B}|} \{\hat{x}_{\hat{i}, \hat{j}}=\operatorname*{argmax}\limits_{x_{i, j}} {x_{i, j} r_{i}}\}$, which represents the $|\mathscr{B}|$ UEs with the maximum data rate requirement among all UEs. Thus, the objective value of $\Lambda_{3}$ should be equal or bigger than $\Phi_{2}(\cup_{\hat{i}=y}^{y-1+|\mathcal{B}|} {\hat{x}_{\hat{i}, \hat{j}}^{'}})$. Then, $\Phi_{2}(\cup_{\hat{i}=y}^{y-1+|\mathcal{B}|} {\hat{x}_{\hat{i}, \hat{j}}^{'}}) \leq \Phi_{2}(\Lambda_{3})$, and $\epsilon_{\hat{i}}\Phi_{2}(\cup_{\hat{i}=y}^{y-1+|\mathcal{B}|} {\hat{x}_{\hat{i}, \hat{j}}^{'}}) < \Phi_{2}(\Lambda_{3})$. Therefore, we have $\Phi_{3}(\bar{x}_{i, j}^{*}) < \Phi_{2}(\cup_{\tilde{i}=1}^{y-1}{\tilde{x}_{\tilde{i}, \tilde{j}}}) + \Phi_{2}(\Lambda_{3})$ and $\Phi_{3}(\bar{x}_{i, j}^{*}) < \Phi_{2}(\Lambda_{1})+ \Phi_{2}(\Lambda_{3})$, implying that the objective values of set $\Lambda_{1}$ and $\Lambda_{3}$ are bigger than that of $\Phi_{3}(\bar{x}_{i, j}^{*})$. Meanwhile, the objective value of the problem $\mathscr{P}_{2}$ is smaller or equal to that of problem $\mathscr{P}_{3}$, $\Phi_{2}(x_{i, j}^{*}) \leq \Phi_{3}(\bar{x}_{i, j}^{*})$. We have $\Phi_{2}(x_{i, j}^{*}) < \Phi_{2}(\Lambda_{1})+ \Phi_{2}(\Lambda_{3})$, either $\Phi_{2}(\Lambda_{1}) \geq \frac{1}{2}\Phi_{2}(x_{i, j}^{*})$ or $\Phi_{2}(\Lambda_{3}) \geq \frac{1}{2}\Phi_{2}(x_{i, j}^{*})$ . Thus, $\max(\Phi_{2}(\tilde{x}_{\tilde{i}, \tilde{j}}), \Phi_{2}(\hat{x}_{\hat{i}, \hat{j}})) \geq \frac{1}{2}\Phi_{2}(x_{i, j}^{*})$, which means that the lower bound of the AA-UPB algorithm is bigger than $\frac{1}{2}$ of the optimal value of problem $\mathscr{P}_{2}$ and the approximation ratio of the AA-UPB algorithm is $\frac{1}{2}$.
\end{proof}

\vspace{-4mm}
\begin{align}
\mathscr{P}_{4}:&\max_{\gamma_{j}, h_{j}}\quad  \sum_{i}\sum_{j}{x_{i, j} r_{i}}  \nonumber\\
s.t.:\nonumber & \\
&C1: \gamma_{j} \in \mathscr{V}_{1},\quad\forall j \in \widetilde{\mathscr{B}},                                     \nonumber\\
&C2: h_{j} \in \mathscr{V}_{2}, \quad\forall j \in \widetilde{\mathscr{B}}.                                       \label{eq:e12}
\end{align}

\vspace{-2mm}
\subsection{Solving the DBS placement Problem}\label{DBSP-problem}
The UE association, power and bandwidth allocation are determined in the last subsection. Here, we try to find the best positions to place all DBS which can maximize the total throughput of the network. Problem $\mathscr{P}_{0}$, given $\tilde{x}_{i, j}$,  $\tilde{p}_{i, j}$ and $\tilde{b}_{i, j}$, can be transformed into problem $\mathscr{P}_{4}$. We propose an optimal DBS placement algorithm (\emph{Opt-DBS-Placement}), which utilizes the exhaustive search method \cite{3D_UAV_placment:2018IEEE_WCL} to solve the problem $\mathscr{P}_{4}$, as depicted in Algorithm~\ref{opt-DBSP}. Here, $\left. \Phi_{4}(\gamma_{j}, h_{j})= \Phi \right |_{x_{j}=\tilde{x}_{j}, p_{i, j}=\tilde{p}_{i, j}, b_{i, j}=\tilde{b}_{i, j}}$ is the objective function of $\mathscr{P}_{4}$.

\begin{algorithm}
\caption{The optimal DBS placement algorithm (\emph{Opt-DBS-Placement})} \label{opt-DBSP}
\small
\SetKwData{Left}{left}\SetKwData{This}{this}\SetKwData{Up}{up}
\SetKwFunction{Union}{Union}\SetKwFunction{FindCompress}{FindCompress}
\SetKwInOut{Input}{Input}\SetKwInOut{Output}{Output}
\Input{$\mathscr{B}$, $\mathscr{U}$, $\mathscr{V}_{1}$ , $\mathscr{V}_{2}$, $\tilde{x}_{i, j}$,  $\tilde{p}_{i, j}$ and $\tilde{b}_{i, j}$\;}
\Output{$\hat{\gamma_{j}}^{*}$ and $\hat{h}_{j}^{*}$\;}
\nl \For{$\hat{\gamma}_{j} \in \mathscr{V}_{1}$}{
\nl \For{$\hat{h}_{j} \in \mathscr{V}_{2}$}{
\nl  update the locations of all DBSs $(\hat{\gamma}_{j}, \hat{h}_{j})$\;
\nl  update $\tilde{x}_{i, j}$,  $\tilde{p}_{i, j}$ and $\tilde{b}_{i, j}$\;
\nl  obtain the objective value, $\Phi_{4}(\hat{\gamma}_{j}, \hat{h}_{j})$\;}}
\nl  calculate $(\hat{\gamma}_{j}^{*}, \hat{h}_{j}^{*})=\operatorname*{argmax}\limits_{\hat{\gamma}_{j}, \hat{h}_{j}} {\Phi_{4}(\hat{\gamma}_{j}, \hat{h}_{j})}$\;
\nl  return $\hat{\gamma}_{j}^{*}$, $\hat{h}_{j}^{*}$.
\end{algorithm}

\begin{theorem}
The Opt-DBS-Placement algorithm produces the optimal positions of all DBSs in the horizontal and vertical dimensions.
\end{theorem}\label{th2}
\vspace{-2mm}
\begin{proof}
Since $\Phi_{4}(\gamma_{j}, h_{j})$ is the objective value of $\mathscr{P}_{4}$, $\Phi_{4}(\hat{\gamma}_{j}, \hat{h}_{j})$  is the total throughput of the network for given locations of all DBSs in the horizontal and vertical dimensions ($\hat{\gamma}_{j}$ and $\hat{h}_{j}$), and determined UE association power and bandwidth assignment ($\tilde{x}_{j}$, $\tilde{p}_{i, j}$ and $\tilde{b}_{i, j}$). Meanwhile, $\left. \Phi_{4}(\hat{\gamma}_{j}^{*}, \hat{h}_{j}^{*})= \Phi \right |_{\gamma_{j}=\hat{\tau}_{j}^{*}, h_{j}=\hat{h}_{j}^{*}}= \max \limits_{\hat{\gamma}_{j}, \hat{h}_{j}} {\Phi_{4}(\hat{\gamma}_{j}, \hat{h}_{j})}$, $(\hat{\gamma}_{j}^{*}, \hat{h}_{j}^{*})=\operatorname*{argmax}\limits_{\hat{\gamma}_{j}, \hat{h}_{j}} {\Phi_{4}(\hat{\gamma}_{j}, \hat{h}_{j})}$, Algorithm~\ref{opt-DBSP} has checked all candidate horizontal and vertical positions. Thus, the optimal horizontal and vertical positions are achieved by Algorithm~\ref{opt-DBSP}.
\end{proof}

\vspace{-2mm}
\subsection{Solving the BUD Problem}\label{BUD-problem}
Here, we propose an approximation algorithm based on Algorithm~\ref{AA-UPB} and Algorithm~\ref{opt-DBSP}, which is named Approximation Algorithm for the BUD problem (\emph{AA-BUD}) to solve problem $\mathscr{P}_{0}$, as depicted in Algorithm~\ref{AA-BUD}.

\begin{algorithm}
\small
\caption{Approximation Algorithm for the BUD problem (\emph{AA-BUD})} \label{AA-BUD}
\SetKwData{Left}{left}\SetKwData{This}{this}\SetKwData{Up}{up}
\SetKwFunction{Union}{Union}\SetKwFunction{FindCompress}{FindCompress}
\SetKwInOut{Input}{Input}\SetKwInOut{Output}{Output}
\Input{$\mathscr{B}$, $\mathscr{U}$, $f_{j}^{max}$, $\kappa_{j}$, $r_{i}$, $\mathscr{V}_{1}$ and $\mathscr{V}_{2}$\;}
\Output{$\tilde{x}_{i, j}$, $\tilde{b}_{i, j}$, $\tilde{p}_{i, j}$, $\hat{\gamma}_{j}$ and $\hat{h}_{j}$\;}
\nl \For{$\tilde{\tau}_{j} \in \Lambda_{1}$}{
\nl \For{$\tilde{h}_{j} \in \Lambda_{2}$}{
\nl  update the locations of all DBSs $(\hat{\gamma}_{j}, \hat{h}_{j})$\;
\nl  obtain $\max(\Phi_{2}(\tilde{x}_{\tilde{i}, \tilde{j}}), \Phi_{2}(\hat{x}_{\hat{i}, \hat{j}}))$ by \emph{Algorithm}~\ref{AA-UPB}\;
\nl  update $\tilde{x}_{i, j}$,  $\tilde{p}_{i, j}$ and $\tilde{b}_{i, j}$\;}}
\nl  obtain $\Phi_{4}(\hat{\gamma}_{j}, \hat{h}_{j})$\;
\nl  compute $(\hat{\gamma}_{j}^{*}, \hat{h}_{j}^{*})=\operatorname*{argmax}\limits_{\hat{\gamma}_{j}, \hat{h}_{j}} {\Phi_{4}(\hat{\gamma}_{j}, \hat{h}_{j})}$\;
\nl  calculate $\tilde{x}_{i, j}$, $\tilde{p}_{i, j}$ and $\tilde{b}_{i, j}$.
\end{algorithm}

\vspace{-4mm}
\begin{theorem}
The AA-BUD algorithm is a $\frac{1}{2}$-approximation algorithm of problem $\mathscr{P}_{0}$.
\end{theorem}\label{th3}

\begin{table}[!htb]
\begin{center}
\centering
\footnotesize
\caption{\small Parameters for Simulations}\label{tab:parameters}
\begin{tabular}{{|l|p{90pt}|}}
\hline
$|\mathscr{B}|$                               & $4$ BSs (including $3$ DBSs)   \\
\hline
coverage area of the MBS                      & $1000m \times 1000m$                       \\
\hline
$f_{0}$                                       & 2 GHz                               \\
\hline
$P_{j}^{max}$, $\forall j\in \mathcal{B}$     & $1$ W                        \\
\hline
$P_{j}^{max}$, $\forall j\in \mathcal{B}$     & $1$ W                        \\
\hline
$|\mathscr{U}|$                               & $\{100, 110,  \cdots, 170\}$     \\
\hline
$(a, b, \zeta^{L}, \zeta^{N})$                & $(4.88, 0.43, 0.1, 21)$~\cite{3D_placement_ICC_2016}  \\
\hline
path loss between a UE and the MBS            & $136.8 + 39.1 log10(d_{i,j})$,                     \\
                                              & \quad\quad\quad $d_{i,j}$ in $km$\cite{HetNet:2017:IEEE_TWC}\\
\hline
Rayleigh fading between a UE and              & -8 dB~\cite{FD_Relay_fading:2016:IEEE_ICC}      \\
the MBS                                       &                                                  \\
\hline
$|\mathscr{V}_{1}|$                           & $36$          \\
\hline
$\mathscr{V}_{2}$                             & $\{100, 120, \cdots, 300\}$ m     \\
\hline
$N_{0}$                                       & $- 174$ dBm/Hz                     \\
\hline
$\tau_{0}$                                    & $15$ kHz                      \\
\hline
$\tau^{SI}$                                   & $130$ dB~\cite{SI_cancellation_130db_2013}    \\
\hline
$r_{i}$                                       & $\{0.5, 1, 1.5, 2\}$ Mbps        \\
\hline
$f^{max}$                                     & $1200$ SCs                \\
\hline
$f_{j}^{max}$                                 & $300$            \\
\hline
\end{tabular}
\end{center}
\vspace{-4mm}
\end{table}

\vspace{-2mm}
\begin{proof}
It is easy to conclude \emph{Theorem} $3$ from \emph{Theorem} $1$ and \emph{Theorem} $2$. In other words, the lower bound of Algorithm~\ref{AA-BUD} is bigger than $\frac{1}{2}$ of the optimal value of problem $\mathscr{P}_{2}$ and the approximation ratio of the AA-BUD algorithm is $\frac{1}{2}$.
\end{proof}

\section{Performance Evaluation} \label{sec:evaluation}


MATLAB is used to run the simulations \cite{Liang_DC:2018:IEEE_TGCN}, and we run each simulations $200$ times to achieve average results. The maximum transmission power of a DBS is set as $40$ dBm, and that of a UE is set as $23$ dBm. We assume there are three DBSs in the network ($|\widetilde{\mathscr{B}}|=3$), and all DBS are placed at the same altitude. The locations of UEs are generated through a Mat{\'e}rn cluster process \cite{Liang_DBS:2018_WCL}. The simulation parameters are summarized in Table~\ref{tab:parameters}.

\begin{figure}[!htb]
    \centering
    \includegraphics[width=0.9\columnwidth]{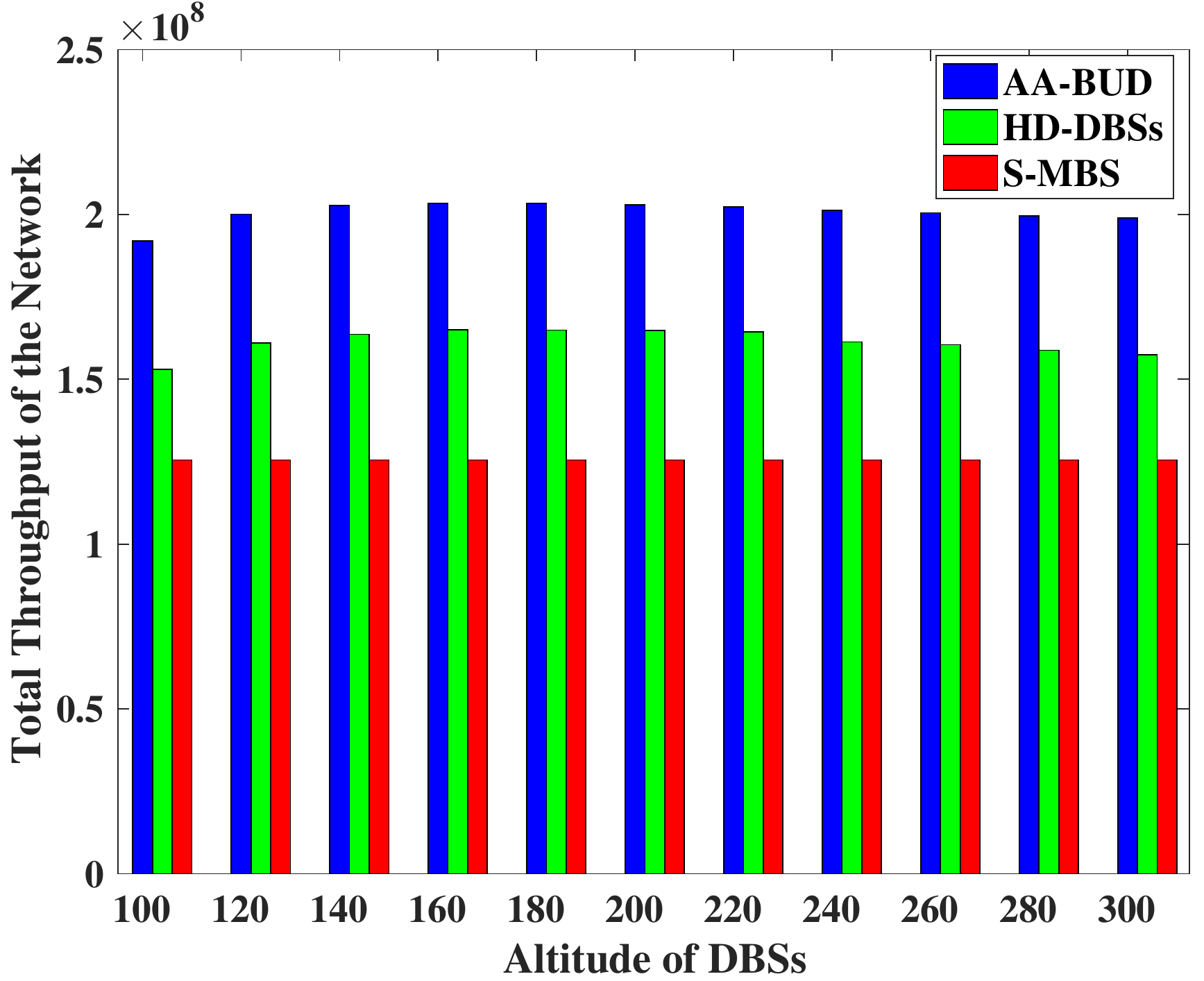}
    \caption{\small Total throughput versus altitude with $170$ UEs.} \label{fig:GC_altitude}
\end{figure}

\begin{figure}[!htb]
    \centering
    \includegraphics[width=0.9\columnwidth]{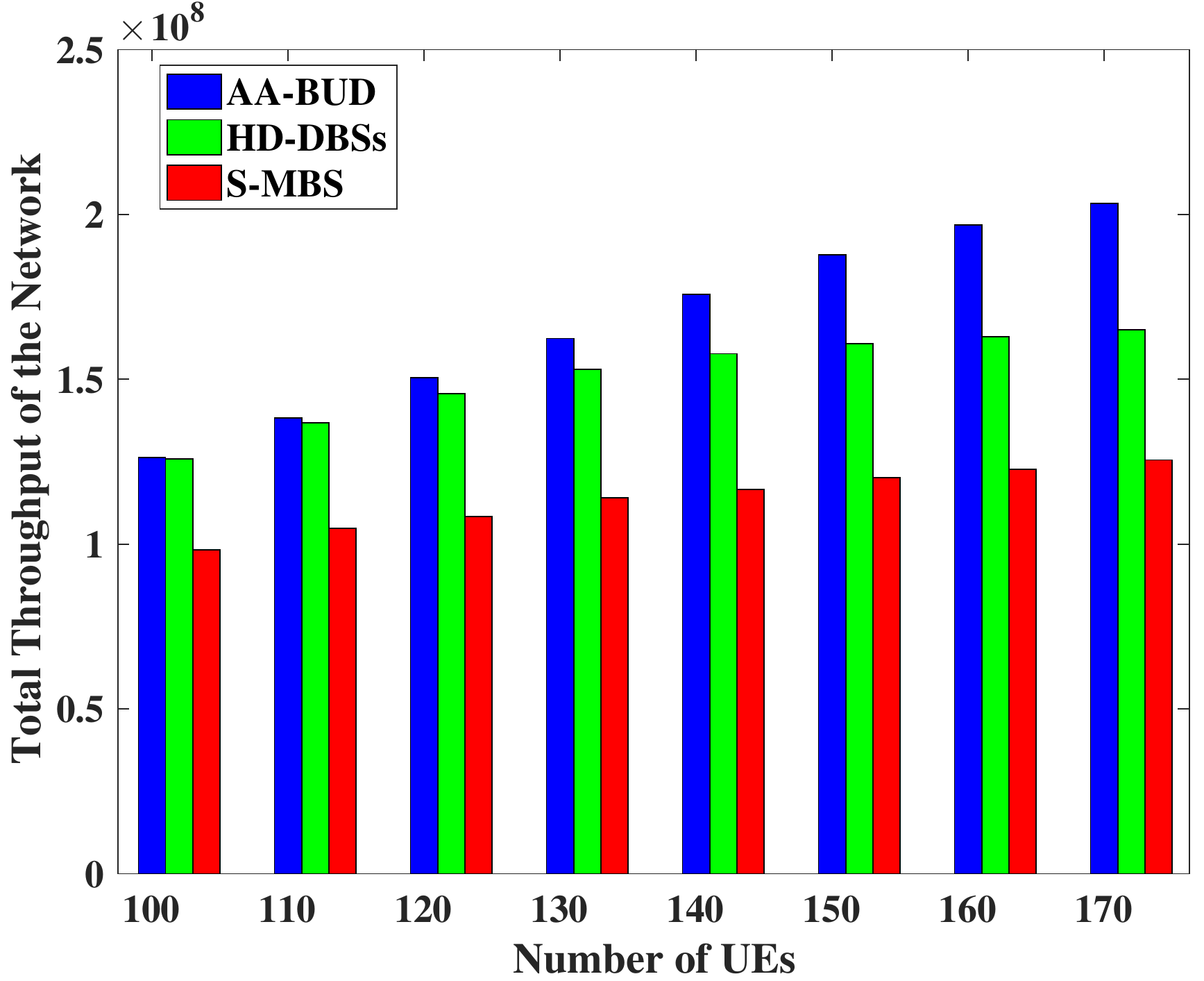}
    \caption{\small Total throughput versus the number of UEs at $160$m altitude.} \label{fig:GC_workload}
    \vspace{-2mm}
\end{figure}

We evaluate the performance of the AA-BUD algorithm with two baseline algorithms. One is the single MBS algorithm without any DBSs (\emph{S-MBS}), and the other algorithm named \emph{HD-DBSs} with half-duplex enabled DBSs. The HD-DBSs algorithm utilizes the same DBS placement strategy, UE association strategy, and the same power and bandwidth assignment strategy as the AA-UPB algorithm.

\begin{figure}[!htb]
    \centering
    \includegraphics[width=0.9\columnwidth]{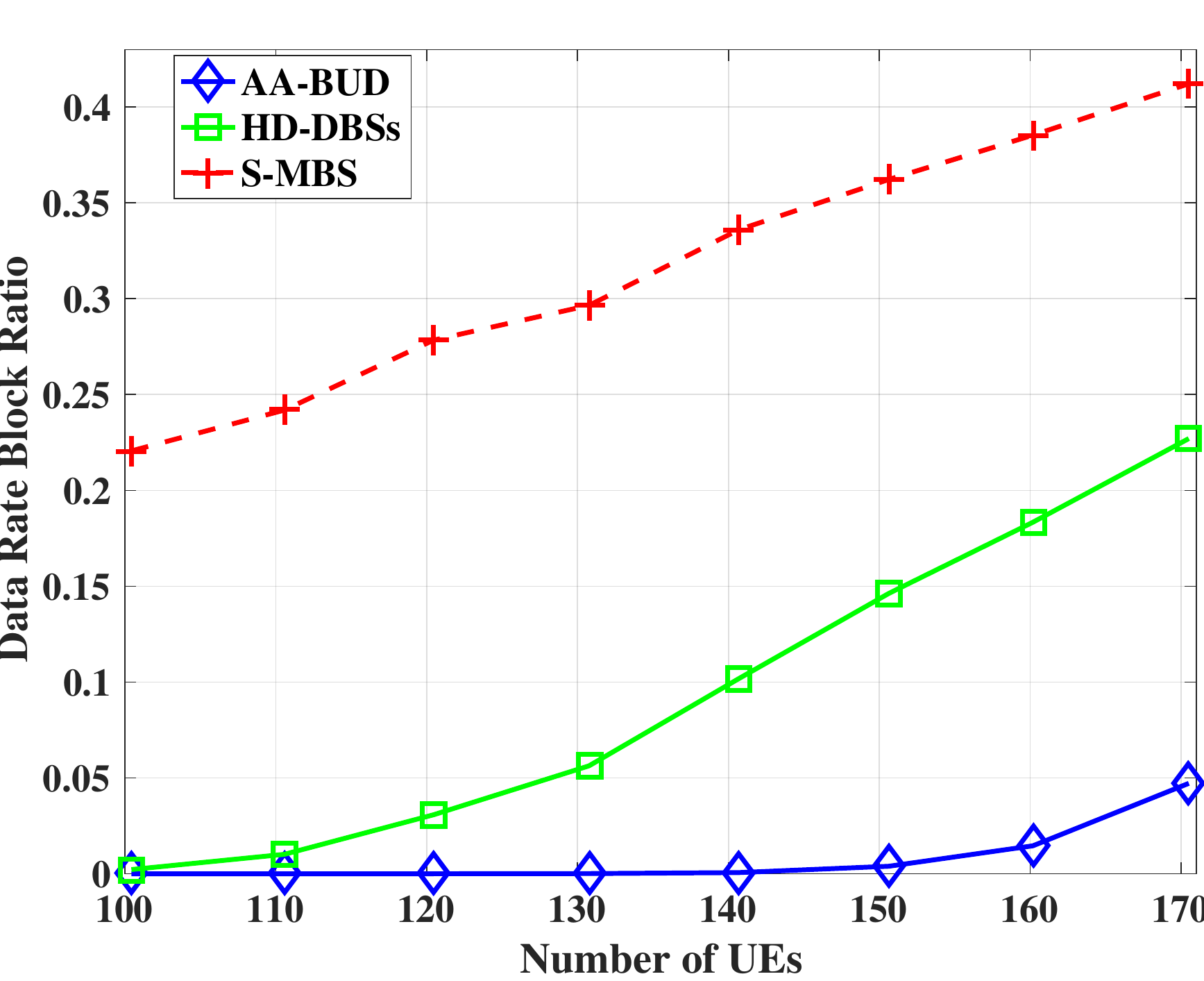}
    \caption{\small Data rate block ratio at $160$m altitude.} \label{fig:GC_block}
    \vspace{-4mm}
\end{figure}

The total throughput performance versus the altitude with $170$ UEs is shown in Fig.~\ref{fig:GC_altitude}. The HD-DBSs algorithm obtains the maximum throughput at $120$m while the AA-BUD algorithm achieves the maximum throughput at $160$m. For the HD-DBSs algorithm, the bottleneck of the uplink communications is the backhaul links (DBS-MBS links) while that of the AA-BUD algorithm is the access links (UE-DBS links or UE-MBS links). This is because the UEs can utilize more frequency spectra when FD-enabled DBSs are operated by the AA-BUD algorithm. For altitude lower than $160m$ of the AA-BUD algorithm, the path loss is dominated by NLoS-path-loss, which decreases as the altitude increases. For altitude higher than $160m$ using the AA-BUD algorithm, the path loss is dominated by LoS-path-loss, which increases as the
altitude increases.

The total throughput results versus the workload with $160$m altitude are shown in Fig.~\ref{fig:GC_workload}. The AA-BUD algorithm achieves up to $23\%$ and $62\%$ improvement of the total throughput as compared to the S-MBS algorithm and HD-DBSs algorithm, respectively. The total throughput of all algorithms increases as the number of UEs increases. This is because all algorithms try to serve UEs with better channel conditions first and then provision the remaining UEs. Hence, less radio resources can be used to provision the same number of UEs but with better channel conditions, and then more UEs can be provisioned by the remaining radio resources.

The data rate block ratio versus workload at $160$m altitude is shown in Fig.~\ref{fig:GC_block}. Here, the data rate block ratio is defined as the data rate requirement of provisioned UEs of the uplink communications over the total uplink data rate requirement of all UEs. Obviously, the AA-BUD algorithm exhibits the best performance with the lowest data rate block ratio, and all UEs are provisioned until the number of UEs reaches $150$. Evaluation results have demonstrated that the AA-BUD algorithm is superior to the baseline algorithms.

\section{Conclusion}\label{sec:conclusion}
In this paper, we have investigated the \underline{b}ackhaul-aware \underline{u}plink communications in a full-duplex \underline{D}BS-aided HetNet (\emph{BUD}) problem with the target to maximize the total throughput of the network for the uplink communications. The DBSs are full-duplex enabled, and the MBS and all UEs are half-duplex enabled. Free space optics (\emph{FSO}) terminals are used to connect the MBS to the core network. The proposed AA-BUD algorithm has been proved to be a $\frac{1}{2}$-approximation algorithm that is capable of obtaining the optimal horizontal and vertical dimensions of DBSs. Evaluation results have also demonstrated that the proposed AA-BUD algorithm is superior to the other baseline algorithms with up to $62\%$ improvement of the uplink throughput.


\clearpage
\bibliographystyle{IEEEtran}

\begin{IEEEbiography}[{\includegraphics[width=1in,height=1.25in,clip,keepaspectratio]{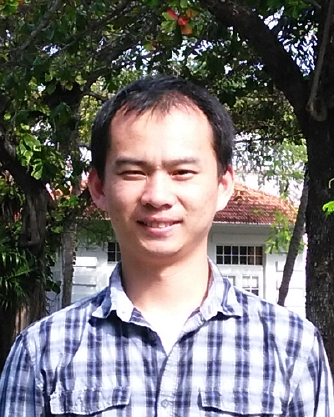}}] {Liang Zhang} [S’15] (lz284@njit.edu) received his M.S. degree from the Department of Information and Communication Engineering, University of Science and Technology of China, in 2014. He is currently pursuing a Ph.D. in ECE at New Jersey Institute of Technology. He received the prize of the National Scholarship of Graduate Students in China in 2013, the Best Paper Award at IEEE ICNC in 2014, and a Travel Grant Award from IEEE GLOBECOM in 2016. His research interests include drone-mounted base-station communications, wireless communications, datacenter networks and optical networks.
\end{IEEEbiography}

\begin{IEEEbiography}[{\includegraphics[width=1in,height=1.25in,clip,keepaspectratio]{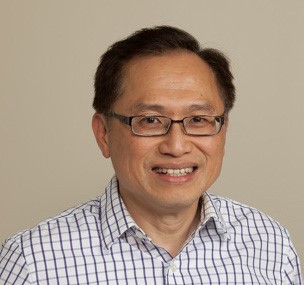}}]{Nirwan Ansari}[S'78-M'83-SM'94-F'09]  (nirwan.ansari@njit.edu) is Distinguished Professor of Electrical and Computer Engineering at the New Jersey Institute of Technology (NJIT). He authored Green Mobile Networks: A Networking Perspective (Wiley-IEEE, 2017) with T. Han, and co-authored two other books. He has also (co-)authored more than 600 technical publications, over 290 published in widely cited journals/magazines. He has guest-edited a number of special issues covering various emerging topics in communications and networking. He has served on the editorial/advisory board of over ten journals including as the Associate Editor-in-Chief of IEEE Wireless Communications. His current research focuses on green communications and networking, cloud computing, drone-assisted networking, and various aspects of broadband networks.
He was elected to serve in the IEEE Communications Society (ComSoc) Board of Governors as a member-at-large, has chaired some ComSoc technical and steering committees, has been serving in many committees such as the IEEE Fellow Committee, and has been actively organizing numerous IEEE International Conferences/Symposia/Workshops. He has frequently been delivering keynote addresses, distinguished lectures, tutorials, and invited talks. Some of his recognitions include several Excellence in Teaching Awards, a few best paper awards, the NCE Excellence in Research Award, several ComSoc TC technical recognition awards, the NJ Inventors Hall of Fame Inventor of the Year Award, the Thomas Alva Edison Patent Award, Purdue University Outstanding Electrical and Computer Engineering Award, NCE 100 Medal, and designation as a COMSOC Distinguished Lecturer. He has also been granted more than 40 U.S. patents.
He received a Ph.D. from Purdue University, an MSEE from the University of Michigan, and a BSEE (summa cum laude with a perfect GPA) from NJIT.
\end{IEEEbiography}
\end{document}